\documentclass{emulateapj}
\usepackage{natbib}
\usepackage{times}

\begin{document}

\shorttitle{A SWIFT RISE IN THE GRB RATE} \shortauthors{KISTLER, Y{\"U}KSEL, BEACOM \& STANEK}

\title{An Unexpectedly Swift Rise in the Gamma-ray Burst Rate}

\author{Matthew D. Kistler\altaffilmark{1,2}, Hasan Y{\"u}ksel\altaffilmark{1,2}, John F. Beacom\altaffilmark{1,2,3}, Krzysztof Z. Stanek\altaffilmark{3,2}}

\altaffiltext{1}{Dept.\ of Physics, The Ohio State University, 191 W.\ Woodruff Ave., Columbus, OH 43210}
\altaffiltext{2}{Center for Cosmology and Astro-Particle Physics, The Ohio State University, 191 W.\ Woodruff Ave., Columbus, OH 43210}
\altaffiltext{3}{Dept.\ of Astronomy, The Ohio State University, 140 W.\ 18th Ave., Columbus, OH 43210}


\begin{abstract}
The association of long gamma-ray bursts with supernovae naturally suggests that the cosmic GRB rate should trace the star formation history.  Finding otherwise would provide important clues concerning these rare, curious phenomena.  Using a new estimate of \textit{Swift} GRB energetics to construct a sample of 36 luminous GRBs with redshifts in the range $z$$\,=\,$0$\,-\,$4, we find evidence of enhanced evolution in the GRB rate, with $\sim\,$4 times as many GRBs observed at $z$$\,\approx\,$4 than expected from star formation measurements.  This direct and empirical demonstration of needed additional evolution is a new result.  It is consistent with theoretical expectations from metallicity effects, but other causes remain possible, and we consider them systematically.
\end{abstract}


\section{Introduction}
Long gamma-ray bursts are perhaps the grandest spectacles in astrophysics.  Their connection with core-collapse supernovae \citep{Stanek:2003tw,Hjorth} indicates progenitors that are very massive, short-lived stars, leading one to expect the cosmic GRB history to follow the star formation history (SFH) \citep{Totani,Wijers:1998,Lamb,Blain:2000,Porciani}.  Any deviation from this expectation would provide new information about why a star should die as a GRB, complementing microphysical investigations (see \citet{Meszaros} for a review).

In just the past few years, \textit{Swift} \citep{Gehrels:2004am} and a worldwide network of observers have detected gamma-ray bursts from higher redshifts than was previously possible, sparking renewed interest in the GRB redshift distribution \citep{Berger:2005,Natarajan:2005,Bromm:2006,Daigne,Jakobsson,Le:2006pt,Yuksel:2006qb,Salvaterra,Liang,Guetta:2007,Chary}.  The \textit{Swift} bursts with known redshifts now provide a sufficiently large sample upon which reasonable cuts can be made.  Additionally, improved star formation measurements, compiled in the analysis of \citet{Hopkins:2006bw}, now provide a well-defined baseline for comparison.  Using a simple, model-independent test, we find that the GRB rate history appears to evolve more strongly than the SFH until at least $z\approx4$.  We emphasize that our result establishes the existence of an evolutionary trend, but that, at present, we cannot identify the underlying reason (i.e., the distinction between correlation and causality).  We discuss possible causes, including observational effects, the suggestion that the lower metallicity of the high-$z$ universe could allow more bursts, and other possibilities that may result in additional GRB progenitors.

\begin{figure}[b!]
\includegraphics[width=3.0in,clip=true]{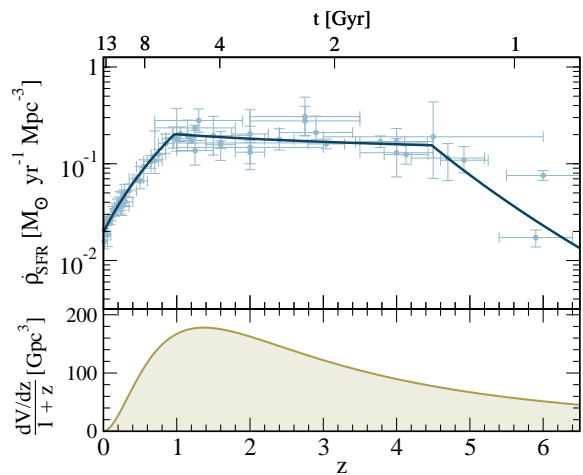}
\caption{\textit{Top panel}: The history of star formation.  Shown are the collection of data (circles) and fit of \citet{Hopkins:2006bw}.
\textit{Bottom panel:} The ``volumetric factor'' in Eq.~(\ref{zdisteq}), which distorts the observed GRB distribution when viewed in $z$.
\label{SFH}}
\end{figure}


\section{The expected GRB rate}
An ever-increasing amount of data has brought about a clearer picture of the history of cosmic star formation.  As shown in Fig.~\ref{SFH}, after a sharp rise up to $z\approx1$, the SFH is nearly flat until $z\approx 4$ with relatively small uncertainties.  These measurements are well-fit by a simple piecewise power law parametrization \citep{Hopkins:2006bw},
\begin{eqnarray}
\dot{\rho}_{\rm SFH}(z) \propto
\begin{array}{lcl}
(1 + z)^{3.44}  		& {\rm :}   	&  z   < 0.97 \nonumber\\
(1 + z)^{-0.26}			& {\rm :}   	&  0.97   < z  < 4.48 \nonumber\\
(1 + z)^{-7.8}  	  & {\rm :}   	&  4.48 < z \,,\nonumber
\end{array}\nonumber
\end{eqnarray}
scaled to $\dot{\rho}_{\rm SF}(0)=0.0197\,M_\odot$~yr$^{-1}$~Mpc$^{-3}$ (a rate per \textit{comoving} volume).  We parametrize the intrinsic (comoving) GRB rate relative to the SFH as $\dot{n}_{\rm GRB}(z) = \mathcal{E}(z)\times \dot{\rho}_{\rm SF}(z)$.  The fraction of bursts that can be seen at a given $z$, 0$\,<\,$$F(z)$$\,<\,$1, depends on the ability to detect the initial burst of gamma rays and to obtain a redshift from the optical afterglow.  We cast the distribution of \textit{observable} GRBs as 
\begin{equation}
	\frac{d\dot{N}}{dz} = F(z) 
	\frac{\mathcal{E}(z)\, \dot{\rho}_{\rm SF}(z)}{\left\langle f_{\rm beam}\right\rangle}
	\frac{dV/dz}{1+z}\,.
	\label{zdisteq}
\end{equation}
This being an observed rate, cosmological time dilation requires the $(1+z)^{-1}$.  The comoving volume element (in terms of the comoving distance, $d_c$), $dV/dz$$\,=\,$$4 \pi\, (c/H_0)\,d_c^2(z)/\sqrt{(1+z)^3\,\Omega_{\rm m}+\Omega_\Lambda}$, peaks at $z\sim2.5$.  Dividing $dV/dz$ by the $1+z$ term yields a \textit{volumetric factor} that peaks at $z\sim1.4$ (see Fig.~\ref{SFH})\footnote{Using $\Omega_\Lambda$$\,=\,$0.7, $\Omega_{\rm m}$$\,=\,$0.3, and $H_0$$\,=\,$70~km/s/Mpc; changing these requires correcting the SFH (see \citet{Hopkins})}.  For a constant $F(z)$, relatively fewer bursts should be observed at $z\sim4$ than $z\sim1$.  GRBs that are unobservable due to beaming are accounted for through $\left\langle f_{\rm beam}\right\rangle$ (see \citet{Bloombeam,Guettabeam,Panaitescu,Kocevski,Nava}).

\begin{figure}[t!]
\includegraphics[width=3.25in,clip=true]{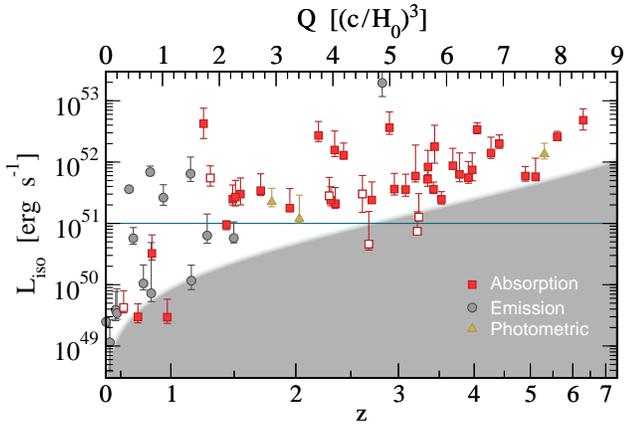}
\caption{The luminosities ($L_{\rm iso}$) of 63 \textit{Swift} gamma-ray bursts, determined from the data of \citet{Butler:2007hw}, vs. z (two $z$$\,<\,$0.1 GRBs are below 10$^{48}$~erg~s$^{-1}$).  Points are plotted linearly in $Q(z)$, as given in Eq.~(\ref{volu}), to account for the volumetric factor.  Open symbols may have underestimated $L_{\rm iso}$.  The shaded region approximates the effective \textit{Swift} detection threshold.  Redshifts were measured as denoted, with medians of $z$$\,=\,$0.8 for emission and $z$$\,=\,$2.9 for absorption.
\label{LisoV1}}
\end{figure}

\section{The Swift observations}
%
%
\textit{Swift} has enabled observers to extend the reach of GRB observations greatly compared to pre-\textit{Swift} times, resulting in a rich data set.  We consider bursts from the \textit{Swift} archive\footnote{http://swift.gsfc.nasa.gov/docs/swift/archive/grb\_table} up to 2007 May 15 with reliable redshifts and durations exceeding $T_{90}$$\,>\,$2~sec.  We estimate each GRB luminosity as $L_{\rm iso}=E_{\rm iso}/[T_{90}/(1+z)]$, where $E_{\rm iso}$ is the isotropic equivalent (beaming-uncorrected) 1$\,-\,$10$^4$~keV rest-frame energy release and $T_{90}$ is the interval observed to contain 90\% of the prompt GRB emission.  To form a uniform $L_{\rm iso}$ set, as displayed in Fig.~\ref{LisoV1}, we use the $E_{\rm iso}$ and $T_{90}$ values given by \citet{Butler:2007hw}.  Note that $L_{\rm iso}$ for several bursts (open symbols) may be underestimated due to $T_{90}$ values that overestimate the GRB duration.

To make the visual density of GRBs in Fig.~\ref{LisoV1} more meaningful, we define a linear $x$-coordinate that ``corrects'' for the effects of the volumetric factor as
\begin{equation}
 Q(z)=\int_{0}^{z} dz^\prime\, \frac{dV/dz^\prime}{1+z^\prime}\,,
 \label{volu}
\end{equation}
which we will quote in terms of $(c/H_0)^3$$\,\approx\,$79~Gpc$^3$.  Removing the influence of the volumetric factor allows for a better ``by-eye'' view, since $d\dot{N}/dQ= (d\dot{N}/dz)/(dQ/dz) =F(Q)\,\dot{n}_{\rm GRB}(Q)/\left\langle f_{\rm beam}\right\rangle$, so that a flat GRB rate would appear as a constant density of GRBs with similar $L_{\rm iso}$ per linear interval in $Q$.

The \textit{Swift} trigger is quite complex, working in the 15$\,-\,$150~keV band and using time-dependent background subtraction and variable time windows in order to maximize burst detection \citep{Gehrels:2004am,Band:2006fj}.  While this sensitivity is ``very difficult if not impossible'' \citep{Band:2006fj} to parametrize exactly, an effective luminosity threshold appears to present in the data (roughly estimated as $\propto\,$$d_l^2$ in Fig.~\ref{LisoV1}).

\begin{figure}[t!]
\includegraphics[width=3.25in,clip=true]{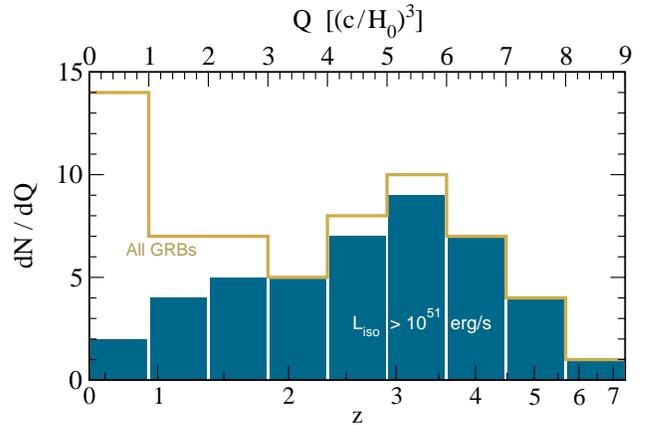}
\caption{The differential GRB distribution versus $Q$.  Outlined bins contain the set of all 63 GRBs (median $z$$\,=\,$2.3), while shaded bins contain the 44 bursts with $L_{\rm iso}$$\,>\,$10$^{51}$~erg~s$^{-1}$ (with a median of $z$$\,=\,$2.9).  The rise seen suggests that GRB rate evolves more strongly than star formation.\\
\label{LisoV2}}
\end{figure}

A representative sample of bursts in a given redshift range can be selected, while avoiding detailed assumptions concerning this threshold, by simply choosing a lower cutoff in $L_{\rm iso}$.  Considering only bursts that could have been seen from anywhere within the redshift range examined allows for the \textit{Swift} contribution to the $F(z)$ term to be treated as constant in $z$.  This technique effectively integrates the GRB luminosity function, $dN/dL_{\rm iso}$, above the chosen $L_{\rm iso}$ cut without assuming its functional form.  This reduces the problem to number counts.


\section{Testing for evolution}
\label{sectevo}
Because $\dot{\rho}_{\rm SF}(z)$ is now reasonably well-measured from $z$$\,\approx\,$0$\,-\,$4 (and nearly flat for 1$\,<\,$$z$$\,<\,$4), we consider GRBs in this range for comparison.  Using only bursts with $L_{\rm iso}$$\,>\,$10$^{51}$~erg~s$^{-1}$ creates a set of GRBs with a minimal expected loss of GRBs up to $z$$\,\lesssim\,$4.  Fig.~\ref{LisoV2} displays the differential distribution of these GRBs versus $Q$.  The bursts above this cut (shaded bins) can be compared to the set of all bursts (outlined).  As can clearly be seen, removing the population of lower luminosity bursts that are only observable at low $z$ reveals a distinct rise in the observed number of ``bright'' bursts.  The drop at $z$$\,\gtrsim\,$4 is likely due to the \textit{Swift} threshold and an overall drop in star formation.

\begin{figure}[t]
\includegraphics[width=3.25in,clip=true]{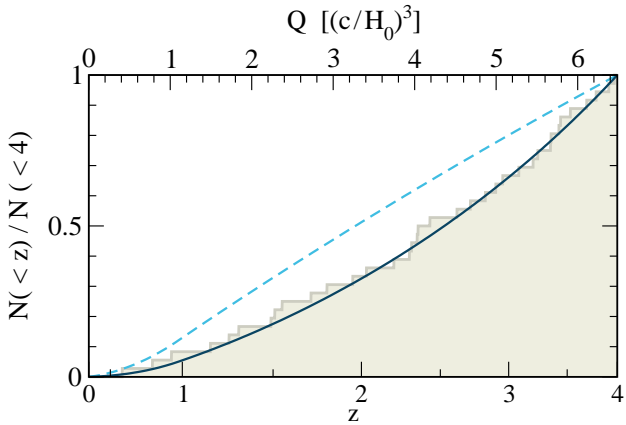}
\caption{The cumulative distribution of the 36 \textit{Swift} GRBs with $L_{\rm iso}$$\,>\,$10$^{51}$~erg~s$^{-1}$ in $z$$\,=\,$0$\,-\,$4 (steps), compared to the expectations from an effective evolution of $(1+z)^{1.5}$ with respect to $\dot{\rho}_{\rm SF}(z)$ (dark solid); and the SFH alone (dashed), which is inconsistent with the data at $\sim 95\%$.\\
\label{KS}}
\end{figure}

To compare the GRB data with the SFH expectation, we make use of Eq.~(\ref{zdisteq}), parameterizing the ``effective evolution'', for simplicity, as $F(z)\times \mathcal{E}(z)\propto (1+z)^\alpha$.  Since the luminosity cut removes the influence of the \textit{Swift} threshold, any $z$-dependence of $F$ would be due to other potential observational effects.  We compare the predicted and observed cumulative GRB distributions in Fig.~\ref{KS}.  This compares the relative trends, independent of the overall normalization and the value of $\left\langle f_{\rm beam}\right\rangle$.  A Kolmogorov-Smirnov test reveals that the SFH fit alone is incompatible at around the $95\%$ level.

Positive evolution results in clear improvement, strengthening indications based upon different analyses and smaller data sets (e.g., \citet{Daigne,Le:2006pt,Yuksel:2006qb}).  We find that the K-S statistic is minimized for $\alpha$$\,=\,$1.5.  To simulate dispersion in the data, we replace the $T_{90}$ values used to calculate $L_{\rm iso}$ with those reported by \textit{Swift} \citep{Sakamoto} and find similar conclusions.  In neither case is any significant correlation of luminosity with $z$ found in this range, disfavoring just an evolving luminosity function \citep{Efron:1998qy}.  Although we have attempted to minimize the loss of GRBs due to the \textit{Swift} threshold, we are more likely to be underestimating burst counts at the upper end of the $z$-range examined, which could imply even stronger evolution.

\section{Is the evolution intrinsic?}
%
%
%
This evolution could result from several causes.  The K-S test disfavors an interpretation as a statistical anomaly.  While $\dot{\rho}_{\rm SF}(z)$ could just be mismeasured, the relatively-small uncertainties over the range in question suggests that this is not the likely origin.  We will discuss in more detail whether it is due to some selection effect (i.e., a changing $F(z)$), or a real physical effect related to changes in the number of progenitors (an evolving $\mathcal{E}(z)$).

It has been suggested that a higher percentage of GRBs may go undetected at low redshifts \citep{Bloom:2003ic,Fiore}.  We perform several simple diagnostics to determine whether the sample of \textit{Swift} GRBs without redshifts contains an overwhelming number of bright, low-$z$ bursts.  To simplify analysis, we consider a sample of bursts that meet the ``detectability'' criteria of \citet{Jakobsson} (in particular, low Galactic extinction and quick \textit{Swift} localization), a set of 50 GRBs with a confirmed $z$ and 47 without.  With this set, using the procedure of Sect.~\ref{sectevo}, the SFH expectation alone is still incompatible at around $95\%$, with $\alpha$$\,=\,$1.5 minimizing the K-S statistic.

On average, more distant bursts are expected to have lower observed gamma-ray fluxes, $\mathcal{F}$, estimated by dividing each 15$\,-\,$350~keV fluence and $T_{90}$ from \citet{Butler:2007hw}.  Indeed, GRBs at $z$$\,>\,$2 have $\left\langle \mathcal{F}_{z>2} \right\rangle \sim 1.4\times 10^{-7}$~erg cm$^{-2}$ s$^{-1}$, compared to $\left\langle \mathcal{F}_{z<2} \right\rangle \sim 2.6\times 10^{-7}$~erg cm$^{-2}$ s$^{-1}$ at $z$$\,<\,$2.  For GRBs without a redshift, the average flux is just $\left\langle \mathcal{F}_{{\rm no}~z} \right\rangle \sim 0.7\times 10^{-7}$~erg cm$^{-2}$ s$^{-1}$.  A two-sample K-S test between the $z$$\,<\,$2 and $z$-less sets reveals that they are incompatible at $\sim$70\%.  Limiting the $z$$\,<\,$2 set only to GRBs with $L_{\rm iso}$$\,>\,$10$^{51}$~erg~s$^{-1}$, this increases to $\sim$99\%.  While not conclusive, these results could be interpreted as most $z$-less bursts being either at high $z$, in which case our evolution may be \textit{underestimated}, or at low $z$ with lower intrinsic luminosities, which may not survive our $L_{\rm iso}$ cut.  Additionally, we examine the fractions of GRBs detected by \textit{Swift}'s UV-Optical Telescope in the \textit{Swift} archive.  Of bursts with a known $z$, 34 were seen by UVOT with $\left\langle z \right\rangle=2.2$, while the 16 not seen have $\left\langle z \right\rangle=3.0$.  Bursts lacking redshifts seem more consistent with the high-$z$ set, with only 8/47 seen by UVOT.

We also consider whether the fraction of observable bursts, $F(z)$$\,=\,$$f_{z}\,f_{Swift}$, is somehow \textit{increasing} with $z$.  While $f_{Swift}$ is difficult to quantify, our selection criteria disfavor incompleteness of our sample at low $z$.  We focus upon the probability of determining a redshift for a given GRB, which we subdivide as $f_{z}$$\,=\,$$f_{\rm E/A}\,f_{\rm AG}\,f_{\rm human}$.  Perhaps the most obvious influence on this term is the fact that at low $z$, most redshifts are determined by observing emission lines (from the host galaxy); while at higher redshifts, nearly every redshift is found through absorption lines in the afterglow spectrum (see Fig.~\ref{LisoV1} and the \textit{Swift} archive).  While this emission/absorption bias, $f_{\rm E/A}$, is not easy to quantify, it should not cause such an evolutionary effect, since most redshifts in our sample are found through absorption.

The observability of an optical afterglow, $f_{\rm AG}$, might be expected to be steeply falling with redshift; however, for a spectrum $\propto t^{-\alpha}\,\nu^{-\beta}$ \citep{Sari:1997qe}, cosmological redshifting may allow the earliest (brightest) portion of the afterglow to be more visible \citep{Ciardi:2000by}.  If the flatness of the obscuration-corrected SFH at moderate $z$ arose from a steeply increasing uncorrected rate and a deceasing dust correction, the detectable afterglow fraction might increase with redshift; this does not appear to be the case \citep{Schiminovich:2004km}.  A paucity of IR detections of hosts of ``dark'' GRBs also argues against a significant obscured fraction \citep{Le Floc'h:2006hu}.  The decision of which telescopes are made available to observe GRBs, along with other such non-intrinsic properties, can be folded into $f_{\rm human}$.  None of these terms appear to be able to increase overall observability with $z$, disfavoring an origin of the trend in $F(z)$.

\section{Potential sources of evolution}
We now investigate whether an evolving $\mathcal{E}(z)$ can explain the observed evolution.  While it is now generally accepted that long GRBs arise from massive stars, the special conditions that are required for such an event are still in question.  In the collapsar model \citep{MacFadyen}, the collapse of a rapidly-rotating, massive stellar core to a black hole powers a jet that is seen as a GRB.  Since every observed supernova coincident with a GRB has been of Type~Ic \citep{Woosley:2006}, the progenitor star should also have lost its outer envelope (without losing precious angular momentum).  Rather strong observational evidence now indicates that GRB host galaxies tend to be faint and metal-poor (e.g., \citet{Fruchter,Stanek:2006gc,Le Floc'h:2003yp,Fynbo:2003sx}), increasing interest in models that use single, low-metallicity stars as a pathway to a collapsar \citep{Yoon:2005tv,Woosley06}.  Decreasing cosmic metallicity may cause the GRB/SN ratio to rise with $z$; e.g., the prediction of \citet{Yoon:2006fr} can be estimated as $(1+z)^{1.4}$ \citep{Yuksel:2006qb}.  While a preference for low-metallicity environments may be the simplest explanation, absorption line metallicity studies remain inconclusive \citep{Savaglio:2006xe,Prochaska}.  One may therefore wonder whether it is possible to concoct an evolutionary scenario without direct progenitor metallicity dependence.

If GRBs are instead produced in binary systems (e.g., \citet{Fryer:1999qs,Podsiadlowski:2004mt,Dale}), some other mechanism might lead to the appearance of evolution.  Since most GRBs appear to occur in star clusters \citep{Fruchter}, where the fraction of massive stars in binaries may be high \citep{Kobulnicky:2006bk}, such channels could be important.  For example, the merger of two $\gtrsim\,$$15\,M_\odot$ stars with $M_1/M_2$$\,\gtrsim\,$0.95 (i.e., ``twins'', which may be common \citep{Pinsonneault:2005qc}) in close orbits ($r\lesssim$~few~AU) can lead to a more-massive, rapidly-rotating core lacking an envelope \citep{Fryer:2004mp}.  In a dense cluster, close binaries tend to end up closer due to gravitational scatterings with interloping stars (Heggie's Law; \citet{Heggie:1975tg}), with a scattering timescale of $\lesssim\,$10~Myr for a stellar density of $\rho \sim 10^6\, M_\odot\, {\rm pc}^{-3}$ \citep{Hut:1992wz}.  An increased rate of ``interloper-catalyzed'' binary mergers (ICBMs) could result from a larger fraction of star formation occurring in such environments at higher $z$, and seen as an enhancement in the GRB rate.  Such a speculative scenario presents a dynamical source of evolution (instead of altering the microphysics) and, as these rates are $\propto\,$$\left\langle \rho^2 \right\rangle$, possibly allow for examination of the ``density contrast'' of star formation.  It is interesting that GRBs were discovered in searches for gamma rays from explosions related to ICBMs of a different sort \citep{Klebesadel:1973iq}.

An evolving IMF, becoming increasingly top-heavy at larger $z$, would increase the relative number of massive stars produced.  Since star formation measurements are primarily based on radiation from such stars, this alone may not lead to apparent evolution, unless the very massive end ($\gtrsim 25 M_\odot$) changed significantly.  Any evidence of evolution in the IMF (e.g., \citet{Wilkins}) provides motivation for considering such a change.  However, this need only occur in those galaxies that host gamma-ray bursts.  A similar effect could result from a larger high-$z$ population of small galaxies resembling low-$z$ GRB host galaxies, seen as evolution in the galaxy luminosity function (e.g., \citet{Ryan}).  An additional possibility is that, if the massive stellar binary fraction decreases with declining metallicity, as might be evidenced in the LMC \citep{Mazeh}, the number of potential single-star progenitors may increase.

In conclusion, the set of \textit{Swift} gamma-ray bursts now allows for model-independent tests of the connection between the GRB and star formation rates.  We present quantitative evidence that
the GRB rate does not simply track star formation over a broad range in redshift; some mechanism, of a presently-unknown nature, is leading to an enhancement in the observed rate of high-$z$ gamma-ray bursts.  The effects of stellar metallicity appear to be a compelling explanation, but cannot be proven yet.  Additional observations will be the only way to discern the root cause of this effect, and allow for proper understanding of the underlying astrophysics.

\acknowledgments
We thank Andrew Hopkins for helpful discussions and Nat Butler, Charles Dermer, and the referee, Virginia Trimble, for useful comments.  We acknowledge use of the \textit{Swift} public data archive.
MDK is supported by the Department of Energy grant DE-FG02-91ER40690; HY and JFB by the National Science Foundation under CAREER Grant PHY-0547102; and all by Ohio State University and CCAPP.
%

\clearpage

\end{document}